\documentclass[runningheads]{llncs}

\usepackage{graphicx}
\usepackage{amsmath}
\usepackage{cite}
\begin{document}

\title{Analysis of Committee Selection Mechanism in Blockchain}

\author{Shiyu Cai}

\maketitle              
\begin{abstract}
The Committee Selection Mechanism can select multiple users of blockchain network to execute a consensus algorithm, such as PBFT. 
In order to guarantee two properties, the mathematical form of the mechanism is relatively limited. Further, if the  mechanism is used in open network, it will bring about an increase in efficiency, but it will reduce the security and practicability of the blockchain network.

\end{abstract}

\section{Introduction}

The rise of Bitcoin\cite{nakamoto2008bitcoin} has drawn many researchers’ attention to look at how to improve cryptocurrencies. Although POW handle the Sybil-attack very well, it reduces the transaction rate, and the blockchain may be forked, resulting in network instability. Subsequent improvements, such as Casper in Ethereum, have brought minor improvements, but have not touched on the roots of the problem. Therefore, people think of whether the traditional Byzantine algorithm can be combined with Blockchain network. On the one hand, the Byzantine algorithm does not need to execute meaningless calculations. On the other hand, if the user can reach a consensus in a certain block every once in a while, then system can avoid forks.

In particular, the traditional Byzantine algorithm cannot be directly applied to open networks for the following reasons: First, the traditional Byzantine algorithm needs to know the number of users in the network in advance, and then set a threshold according to the number of members. Then users accept a value or block when the vote on that value reaches this threshold. In the open network, users come and go freely, and can create new accounts without restrictions. Therefore, it is impossible to obtain the current number of users. Secondly, the number of users of the Byzantine algorithm should not be too much, otherwise the time of consensus will be unacceptable. Finally, all Byzantine algorithms have a strict theoretical constraint, so honest users must be twice as many as malicious users\cite{lamport1982byzantine}. Otherwise, the malicious users can make good people reach a partial consensus on different blocks by deception, which leads to a fork. And in the open network, how to guarantee this constraint is a serious question.

How to the traditional Byzantine algorithm is applied to an open network? Consensus algorithm such as Algorand\cite{gilad2017algorand} provide a novel committee selection mechanism to determine the sub-users that each user use to execute consensus algorithm through a verifiable random function\cite{dodis2005verifiable}. The mechanism has the following characteristics. First, as long as the user publishes the hash value and signature used for generation of sub-users, the other users verify whether the number of sub-users is correct. Second, it determines the number of sub-users by the number of resources of the user. In addition, the number of users with more resources will be relatively more, which will increase the threshold for malicious users to do evil. Third, through the central limit theorem, it is guaranteed that in most cases, among the selected sub-users, the number of honest sub-users and the number of malicious sub-users satisfy certain constraints\cite{chen2016algorand}.

What we are most concerned about is that with what properties, this mechanism can be well applied to the open blockchain network. What is the mathematical form of the mechanism, and whether any probability distribution is competent. Second, given probability distribution, find the lower bound of the ratio of honest users’ resources to total resources. When the ratio is greater than the lower bound, we can guarantee that in most cases, the system will not go wrong. Will this ratio be so high that the practicality and security of this system will decline?

This article is organized in the following manner. In the second chapter, we give some definitions and notations. In the third chapter, we discuss the properties and forms of the probability distribution of the Committee Selection mechanism that satisfies two conditions. In the fourth chapter, we approximate the problem by chernoff inequality and then give an interval . In the fifth chapter, we first explain the impact of the difference caused by the chernoff inequality on the final conclusion is quite small, and then give two  main conclusions.

\section{Notation and Definition}

\subsection{Definition}

\paragraph{User}
An account in blockchain network.

\paragraph{Sub-user}
Sub-users represent the number of members executing consensus algorithm. 
Each sub-user has only one vote. 
A user can have multiple sub-users.

\paragraph{Committee}
The committee is a collection of sub-users.

\paragraph{Honest user}
An honest user is a user that behave correctly according to the algorithm or program.

\paragraph{Malicious user}
A malicious user is a user that behave badly, for example, they can cheat and attack other users.

\paragraph{Resource}
Each accounts has corresponding resources, such as Computing power, money, contribution, space, ranking and etc.
Committee Selection Mechanism select committee member according to how many resources a user has.

\paragraph{Threshold}
Threshold is a value. When the counts of votes on the same block is greater than threshold, a user accept the block.

\subsection{Notation}

\paragraph{$n$}
Total number of users.

\paragraph{$r_i$}
Resources that user i has. For convenience, we assume $r_i\in Z$.

\paragraph{$v_i\sim D(r_i)$}
Random variable of users i, represent how many sub-users of users i in committee. $D(x)$ is distribution of committee selection mechanism with parameter $x$.

\paragraph{$F$}
Acceptable failure probability, usually $10^{-12}$ or $10^{-18}$.

\paragraph{$v_e$}
Excepted value of the number of committee members.

\paragraph{$t_h$}
Threshold of honest sub-users in committee.

\paragraph{$t$}
The ratio of threshold $t_h$ to expected value of the number of committee sub-users.

\paragraph{$R_h(R_m)$}
The total number of resources of honest(malicious) users. 

\paragraph{$c$}
The ratio of honest users' resources to the total resources. 

\paragraph{$R$}
The total number of resources of all users. 

\paragraph{$V_h(V_m)$}
The number of sub-users in committee of honest(malicious) users. 

\paragraph{$V$}
The total number of sub-users in committee of all users.

\section{Property of Committee Selection Mechanism}

In this section, we will discuss what kinds of properties committee selection mechanism in blockchain should hold? We present two properties, sybil-proof and fairness.

\subsection{Assumption}

Firstly, for convenience, we assume all the resources and sub-users is integer. Besides, sub-users is in specific range.
\begin{equation}
    v_i\in Z\cap [0, r_i]
\end{equation}

Secondly, we assume the expected value of sub-users is proportional to the expected number of total sub-users in committee.
\begin{equation} 
E(v_i)\propto v_e
\end{equation}

\subsection{Sybil-proof}

In order to prevent Sybil-attack, committee selection mechanism choose sub-user base on how many resources a users has, such as money, computing power, contribution and etc. For convenience, we assume all the resources is integer. We intuitively analyzed this problem and came to the conclusion that the expected value of sub-users should be proportional to the number of users' resources. 
\begin{equation} 
E(v_i)\propto r_i
\end{equation}
Based on Equation 1 and Equation 2, we can derive the formula for the expected value,
\begin{equation} 
E(v_i)=kv_er_i
\end{equation}
where k represents a constant.

\subsection{Fairness}

Committee Selection Mechanism is fair or not depends on whether the distribution of the sum of the corresponding random variables
generated with small resources is the same as the random variable generate with merged resources.
Let's consider a situation, suppose there is a user with 10 dollars, and two users with 4 dollars and 6 dollars respectively. The random variable $v\sim D(10)$ and the other two random variables $v_1\sim D(4), v_2\sim D(6)$satisfy the following equation,
\begin{equation}
    E(v) = 10cv_e = (6+4)cv_e = E(v_1)+E(v_2)
\end{equation}
Furthermore, if the two are distributed together, it means that no matter how many mall resources the user splits, there will be no negative or positive impact, which is what we mean fairness. An unfair mechanism can lead to greater or lesser risk for people with more resources.

\begin{theorem}
$\forall r_1,r_2\in Z^+, v_1\sim D(r_1), v_2\sim D(r_2)$, if $v_1+v_2\sim D(r_1+r_2)$ then
\begin{equation}
    v_i\sim B(r_i, p)
\end{equation}
where $B(r_i,p)$ refers to Bernoulli test with $r_i$ times and success probability $p$, and $p=\frac{v_e}{R}$
\end{theorem}

\begin{proof}
When the number of resources is 1, the value of v is only 0 and 1 and can be regarded as the Bernoulli distribution $B(1, p)$ with the probability of success being p and the number of tests being 1. Then when the number of resources is $r$, because of the fairness, random variable $v\sim D(r)$ has the same distribution with the sum of $r$ independent random variable $v_i\sim B(1, p)$, which means $v\sim B(r,p)$.

Then the total number of sub-users in committee satisfies the following equation. 
\begin{equation}
    V=\sum v_i \sim B(R,p)
\end{equation}
\begin{equation}
    E(V)=v_e = pR
\end{equation}
then according to equation 8, we can get
\begin{equation}
    p = \frac{v_e}{R}
\end{equation}
\end{proof}

\section{Lower Bound of Ratio of Honest Users' Resource}

In this section, we will discuss a question that is strongly connected to the security of network: 
Given Committee Selection Mechanism and consensus algorithm, how to guarantee that consensus algorithm will execute correctly
with overwhelming probability? 

Consensus algorithm executing correctly means the number of honest sub-users in committee is greater than given threshold,
$V_h > t_h$. Further, any Byzantine fault tolerance algorithm has common constrain, $V_h > 2V_m$. Combing the first two inequality, we can derive an equivalent inequality, $V_h + 2V_m < 2t_h$.

Recall that F is acceptable failure probability. Consensus algorithms running correctly with great probability means
\begin{equation}
    Prob(V_h > t_h) > 1-F
\end{equation}
\begin{equation}
    Prob(V_h +2V_m < 2t_h) > 1-F
\end{equation}

The problem can be described as $c_h\in [0,1]$, try to find max $c_h$, when $R_h > c_h  R$,  
\begin{equation}
    Prob(V_h > t_h) \le F
\end{equation}
\begin{equation}
    Prob(V_h +2V_m < 2t_h) \le F
\end{equation}

We use Chernoff Bounds to estimate the probability\cite{chung2006complex}.We will explain in the following chapters that the difference between the approximation and the exact value is not large. 

We now discuss the first inequality with Chernoff Bounds.
\begin{equation}
    Prob(V_h > t_h) < e^{-\frac{(E(V_h)-t_h)^2}{2E(V_h)}}\le F
\end{equation}
\begin{equation}
    {E(V_h)}^2+2(lnF-t_h)E(V_h)+t_h^2 \ge 0
\end{equation}
where $E(V_h)=pR_h$
\begin{equation}
    \frac{R_h}{R} \ge \frac{t_h}{v_e}+\frac{-lnF+\sqrt{-2t_hlnF+{lnF}^2}}{v_e}
\end{equation}

Then we discuss the second inequality.
\begin{equation}
    Prob(V_h +2V_m < 2t_h) < e^{-\frac{(2t_h-E(V_h+2V_m))^2}{2(E(V_h)+4E(V_m)+\frac{4t_h-2E(V_h+2V_m)}{3})}} \le F
\end{equation}

Record $A$ as $\frac{2lnF}{3v_e}$, and $t$ as $\frac{t_h}{v_e}$, then

\begin{equation}
    \frac{R_h}{R} \ge \frac{-4t-7A+4+\sqrt{49A^2+40At+32A}}{2}
\end{equation}

Then max $c_h$ is equal to the maximum of right value
of formula 16 and formula 18.
\begin{equation}
    c_{approx} = max(\frac{t_h}{v_e}+\frac{-lnF+\sqrt{-2t_hlnF+{lnF}^2}}{v_e}, \frac{-4t-7A+4+\sqrt{49A^2+40At+32A}}{2})
\end{equation}

\section{Conclusion}

In this section, firstly, we will explain that for the final conclusion, the difference between exact value and approximation has quite small impact. 

\subsection{Difference between exact value and approximation}

The exact value $c_{exact} = max\ \{c\}\ s.t.\ \sum_{i=0}^{t_h}{\tbinom{R_h}{i}p^i(1-p)^{R_h-i}} < F$,  
where $R_h = cR$.

Given $F = 10^{-12}, v_e = 4000, t_h = 0.7v_e$, $c_exact \approx 0.797$ and $c_approx \approx 0.8054$.

This difference does not affect the correctness of the conclusion.

\subsection{Main Result}

The main conclusions are two: 
\begin{enumerate}
    \item As shown in figure \ref{fig1}, the ratio of the threshold to the expected value of the committee $t$ is linear with the lower bound of the proportion of honest people to everyone $c$. For example, when t is equal to 0.7, the ratio of honest users' resources to all resources is about 0.8. This may make the network more inclined to become a static network, without good scalability. This conclusion can help network designers understand what percentage of honest users' resources the system will work correctly with overwhelming probability.
    
    \begin{figure}[h]
\centering
\includegraphics[width=0.8\textwidth]{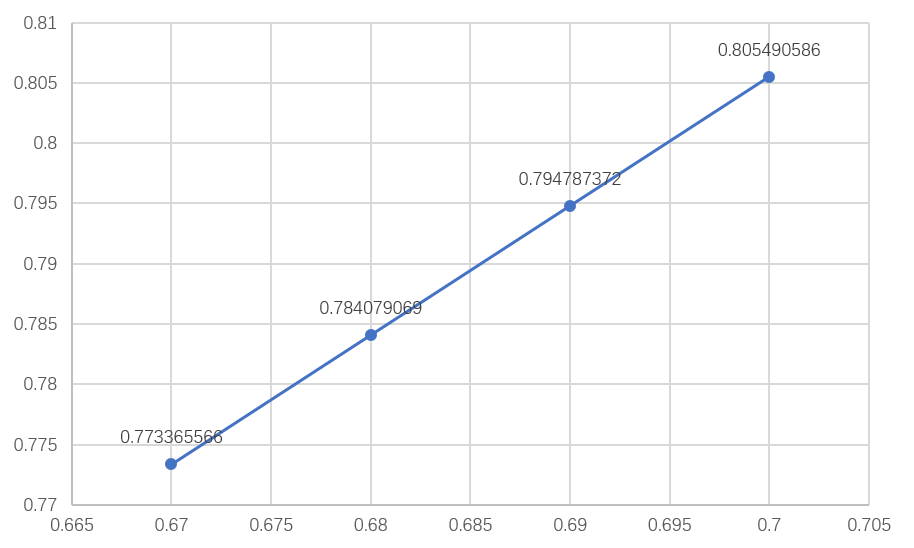}
\caption{The relation between $t$ and $c$}
\label{fig1}
\end{figure}
    
    \item As shown in figure \ref{fig2}, the ratio of malicious users' resource and the probability of acceptable failure are directly exponential. As shown in the figure, when the ratio of malicious users' resources reaches 20\%, the probability of failure is only 10 minus 12 power, and when the bad number reaches 100\%. When it reach 25\%, the probability of failure reached between the $10^{-4}$ and $10^{-5}$. This means that the probability of failure of the system has reached an intolerable level. And for a users' group, when the proportion of the owner's  resources is more than 20 percent, each additional investment of resources, the possibility of doing evil is growing rapidly, which will motivate the group to behave badly.In fact, it is very likely that more than 20\% of the resources will occur. As shown in the figure \ref{fig3}, the $1^{st}$ mine pool on Bitcoin is close to this value, and at some time in the past, it once reached 23\%.
    
    \begin{figure}[h]
\centering
\includegraphics[width=0.8\textwidth]{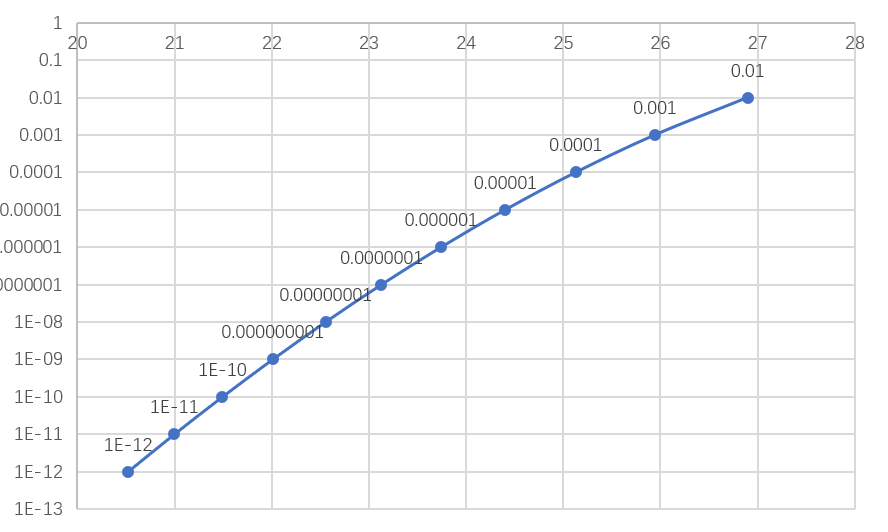}
\caption{The relation between $t$ and $1-c$}
\label{fig2}
\end{figure}

    \begin{figure}
\centering
\includegraphics[width=0.8\textwidth]{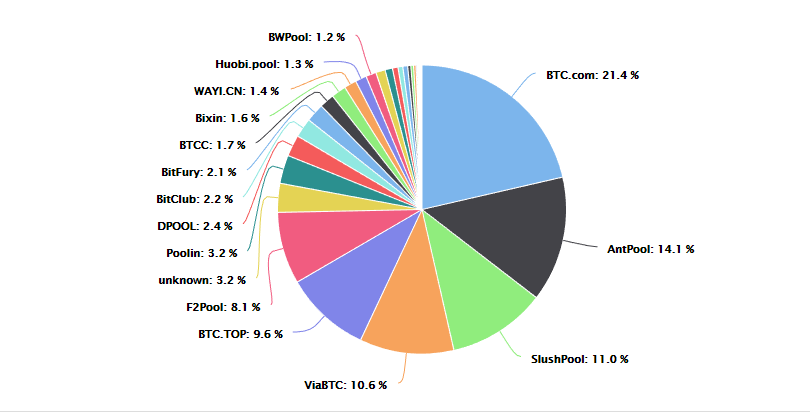}
\caption{The mining power distribution of 2018}
\label{fig3}
\end{figure}

\end{enumerate}

In general, the committee selection mechanism can combine the traditional Byzantine algorithm with the open network. In order to ensure fairness and sybil-proof, the mathematical form of the committee selection mechanism is quite limited. Further, a fair mechanism will pose a full challenge to the security of the network, which is not good news for users of cryptocurrency. Comparing with high transaction efficiency, they may want to invest their money in a safer place.

\bibliography{main.bib}

\begin{thebibliography}{1}
\providecommand{\url}[1]{#1}
\csname url@samestyle\endcsname
\providecommand{\newblock}{\relax}
\providecommand{\bibinfo}[2]{#2}
\providecommand{\BIBentrySTDinterwordspacing}{\spaceskip=0pt\relax}
\providecommand{\BIBentryALTinterwordstretchfactor}{4}
\providecommand{\BIBentryALTinterwordspacing}{\spaceskip=\fontdimen2\font plus
\BIBentryALTinterwordstretchfactor\fontdimen3\font minus
  \fontdimen4\font\relax}
\providecommand{\BIBforeignlanguage}[2]{{%
\expandafter\ifx\csname l@#1\endcsname\relax
\typeout{** WARNING: IEEEtran.bst: No hyphenation pattern has been}%
\typeout{** loaded for the language `#1'. Using the pattern for}%
\typeout{** the default language instead.}%
\else
\language=\csname l@#1\endcsname
\fi
#2}}
\providecommand{\BIBdecl}{\relax}
\BIBdecl

\bibitem{nakamoto2008bitcoin}
S.~Nakamoto \emph{et~al.}, ``Bitcoin: A peer-to-peer electronic cash system,''
  2008.

\bibitem{lamport1982byzantine}
L.~Lamport, R.~Shostak, and M.~Pease, ``The byzantine generals problem,''
  \emph{ACM Transactions on Programming Languages and Systems (TOPLAS)},
  vol.~4, no.~3, pp. 382--401, 1982.

\bibitem{gilad2017algorand}
Y.~Gilad, R.~Hemo, S.~Micali, G.~Vlachos, and N.~Zeldovich, ``Algorand: Scaling
  byzantine agreements for cryptocurrencies,'' in \emph{Proceedings of the 26th
  Symposium on Operating Systems Principles}.\hskip 1em plus 0.5em minus
  0.4em\relax ACM, 2017, pp. 51--68.

\bibitem{dodis2005verifiable}
Y.~Dodis and A.~Yampolskiy, ``A verifiable random function with short proofs
  and keys,'' in \emph{International Workshop on Public Key
  Cryptography}.\hskip 1em plus 0.5em minus 0.4em\relax Springer, 2005, pp.
  416--431.

\bibitem{chen2016algorand}
J.~Chen and S.~Micali, ``Algorand,'' \emph{arXiv preprint arXiv:1607.01341},
  2016.

\bibitem{chung2006complex}
F.~Chung, F.~R. Chung, F.~C. Graham, L.~Lu, K.~F. Chung \emph{et~al.},
  \emph{Complex graphs and networks}.\hskip 1em plus 0.5em minus 0.4em\relax
  American Mathematical Soc., 2006, no. 107.

\end{thebibliography}

\end{document}